\documentclass[%
 reprint,superscriptaddress,
 showpacs,
 amsmath,amssymb,
 aps,
 prb
]{revtex4-1}

\usepackage{graphicx}
\usepackage{dcolumn}
\usepackage{bm}
\usepackage{hyperref}

\begin{document}
\preprint{APS/123-QED}
\title{Low-temperature illumination and annealing of ultra-high quality quantum wells}

\author{M. Samani}
	\affiliation{Quantum Matter Institute, University of British Columbia, Vancouver, British Columbia, V6T1Z4, Canada}
	\affiliation{Department of Physics and Astronomy, University of British Columbia, Vancouver, British Columbia, V6T1Z4, Canada}
\author{A. V. Rossokhaty}
	\affiliation{Quantum Matter Institute, University of British Columbia, Vancouver, British Columbia, V6T1Z4, Canada}
	\affiliation{Department of Physics and Astronomy, University of British Columbia, Vancouver, British Columbia, V6T1Z4, Canada}
\author{E. Sajadi}
	\affiliation{Quantum Matter Institute, University of British Columbia, Vancouver, British Columbia, V6T1Z4, Canada}
	\affiliation{Department of Physics and Astronomy, University of British Columbia, Vancouver, British Columbia, V6T1Z4, Canada}
\author{S. L\"{u}scher}
	\affiliation{Quantum Matter Institute, University of British Columbia, Vancouver, British Columbia, V6T1Z4, Canada}
	\affiliation{Department of Physics and Astronomy, University of British Columbia, Vancouver, British Columbia, V6T1Z4, Canada}
\author{J. A. Folk}
	\email{jfolk@physics.ubc.ca}
	\affiliation{Quantum Matter Institute, University of British Columbia, Vancouver, British Columbia, V6T1Z4, Canada}
	\affiliation{Department of Physics and Astronomy, University of British Columbia, Vancouver, British Columbia, V6T1Z4, Canada}
\author{J. D. Watson}
	\affiliation{Department of Physics and Astronomy, Purdue University, West Lafayette, Indiana, 47907, USA}
	\affiliation{Birck Nanotechnology Center, Purdue University, West Lafayette, Indiana, 47907, USA}
\author{G. C. Gardner}
	\affiliation{Birck Nanotechnology Center, Purdue University, West Lafayette, Indiana, 47907, USA}
    \affiliation{School of Materials Engineering, Purdue University, West Lafayette, Indiana, 47907, USA}
\author{M. J. Manfra}
	\affiliation{Department of Physics and Astronomy, Purdue University, West Lafayette, Indiana, 47907, USA}
	\affiliation{Birck Nanotechnology Center, Purdue University, West Lafayette, Indiana, 47907, USA}
\affiliation{School of Materials Engineering, Purdue University, West Lafayette, Indiana, 47907, USA}
	\affiliation{School of Electrical and Computer Engineering, Purdue University, West Lafayette, Indiana, 47907, USA}

\date{\today}
\begin{abstract}
	The effects of low temperature illumination and annealing on fractional quantum Hall (FQH) characteristics of a GaAs/AlGaAs quantum well are investigated. Illumination alone, below 1 K, decreases the density of the 2DEG electrons by more than an order of magnitude and resets the sample to a repeatable initial state. Subsequent thermal annealing at a few Kelvin restores the original density and dramatically improves FQH characteristics. A reliable illumination and annealing recipe is developed that yields an energy gap of 600 mK for the 5/2 state.
\end{abstract}

\maketitle

The fractional quantum Hall (FQH) state at filling factor $\nu=5/2$ has been the subject of extensive experimental and theoretical scrutiny since its discovery over 25 years ago.\cite{OriginalFiveHalfObs} The $5/2$ state, and its conjugate at $\nu=7/2$, are unique among more than 80 other observed FQH states in that they are the only incompressible states to have been observed at even denominator filling factors in a single layer two-dimensional electron gas (2DEG). A primary reason for such widespread interest in the 5/2 state is the expectation that it may possess non-Abelian quasiparticle statistics.

Like many of the more exotic FQH states, the $\nu=5/2$ state is very fragile. The incompressible state develops fully only in samples with mobility well above $10^6$ cm$^2$/Vs, and only at temperatures below 50 mK. The fragility of the 5/2 state makes experimental measurements very challenging, and has limited progress in understanding its intrinsic microscopic description. The same can be said of other weak FQH states, such as the one at $\nu=12/5$. The 12/5 state may also be non-Abelian\cite{TwelveFifth}, but its activation energy gap is even smaller than that of 5/2 state\cite{TwelveFifthGap}. It is therefore desirable to obtain samples in which FQH states are more robust.

One route to improving FQH activation energy gaps is to optimize the molecular beam epitaxy (MBE) growth recipe.\cite{AlGaAsAlloys,MbeOnly, MBEreview} The highest mobility samples today are GaAs/AlGaAs quantum wells with silicon $\delta$-doping placed in secondary, extremely narrow, quantum wells on both sides of the primary quantum well. These secondary wells are often referred to as ``doping wells''. This strategy makes it easier for charges to equilibrate among the dopants for improved screening. Furthermore, the placement of dopants in separate doping wells---that is, in a GaAs layer as opposed to AlGaAs/AlAs barrier---increases the chance that the dopant energy levels will be shallow, not far below the chemical potential. This increases their ionization fraction dramatically which in turn reduces the disorder potential due to random ionization states. The sample used in the present experiments is of this doping well variety.

Crystal growers have made substantial progress optimizing the quality of the 2DEG samples for FQH characteristics, but it is well known that the growth recipe is not the only parameter that affects the activation energy gap.\cite{AlGaAsAlloys} Illumination during the cooldown process from room temperature to millikelvins also plays an important role, including even the timing of turning the light off. Historically samples have been illuminated down to temperatures around 4.2 K, but what physical parameters set this temperature have not been studied in detail.



In this paper, we investigate a protocol involving illumination and thermal annealing whereby the density, mobility, and FQH quality of a GaAs doping well 2DEG can be optimized after cooldown in a dilution refrigerator. Illumination at millikelvin temperatures is found initially to lower the carrier density of the 2DEG by more than a factor of 30 during the illumination.
Subsequent warming of the 2DEG to temperatures above 2 K causes the density to return to its nominal value, and generates superb FQH characteristics. By optimizing heating parameters, an activation energy gap $\Delta_{5/2}$=600$\pm$10 mK for the $\nu=5/2$ state is obtained, one of the highest values reported.
The outcome is found to be extremely repeatable: the 2DEG can be brought to the same high quality state cooldown after cooldown. It can even be reset {\it in situ} after its mobility and FQH characteristics are severely degraded by an electrostatic shock.

\begin{figure}[htp]
	\begin{center}
	\includegraphics{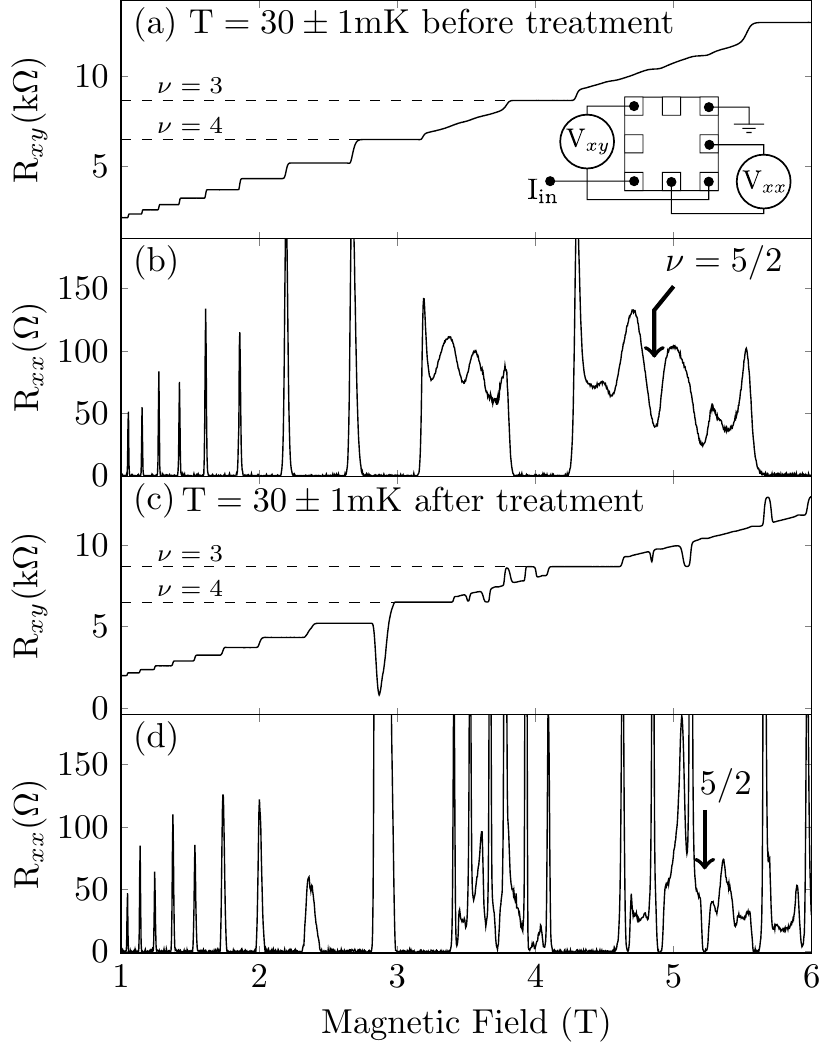}
	\caption{Hall resistance $\mathrm{R}_{xy}$ (a), and longitudinal resistance $\mathrm{R}_{xx}$ (b), up to filling factor $\nu=2$ immediately after cooldown in the dark. (c) $\mathrm{R}_{xy}$ and (d) $\mathrm{R}_{xx}$ after illuminating light on the sample at base temperature for 30 minutes, then annealing the sample at 2.25 K for 20 minutes.\label{beforeaftertreatment}}
	\end{center}
\end{figure}

\begin{figure}[htp]
	\begin{center}
	    \includegraphics{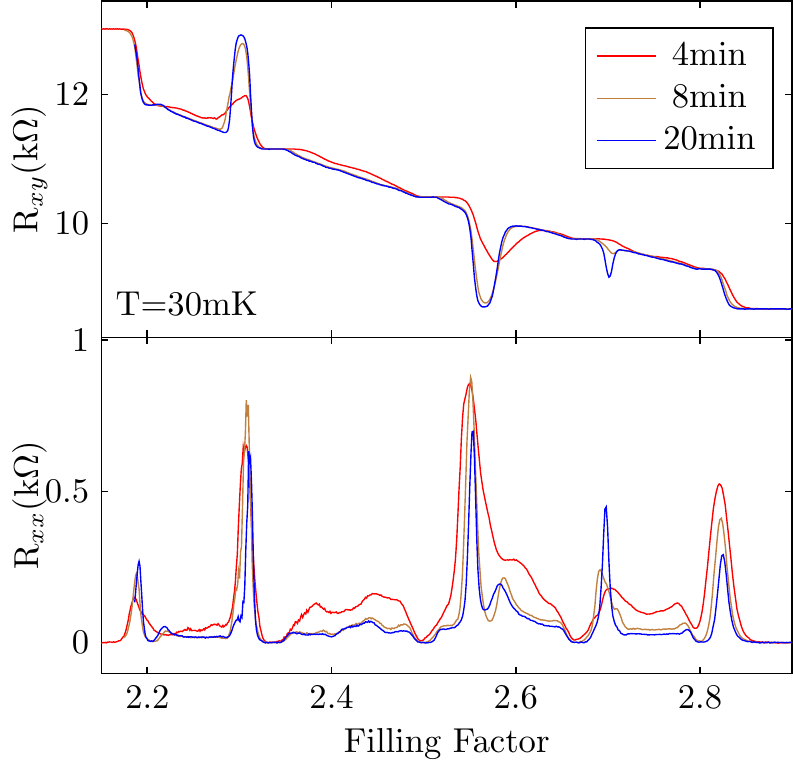}
	    \caption{(Colour online) Longitudinal and Hall resistance after 30 minutes of LED illumination, then 4, 8, and 20 minutes of annealing time at 2.25 K. FQH states are partially developed and reentrant integer quantum Hall states are asymmetric after 4 minutes of annealing, but improve with longer annealing.\label{evolution}}
	\end{center}
\end{figure}

The quantum well under study in this work was grown by the doping well strategy described in Ref.\citenum{MBEreview}, yielding a nominal carrier density of $3.1\times 10^{11}$cm$^{-2}$ and low temperature mobility of $1.5\times 10^7$ cm$^2$/Vs. Eight indium pads were annealed into the $5\times5$ mm$^2$ sample to establish ohmic contact to the 2DEG; the contacts are midway along the edges and at the corners. The sample was thermally sunk to the gold-plated base of a ceramic chip carrier by melted indium. Gold bond wires connected each indium contact to chip carrier contacts, and an additional 14 bond wires connected the carrier base to well-cooled measurement wires. Two resistors mounted to the back side of the chip carrier were used for rapid control of the sample temperature: a metal film resistor (``local heater'') was for heating, while a nearby carbon resistor (``local thermometer'') monitored the temperature.\cite{CarbonRes}    

The sample was cooled in a dilution refrigerator with a base temperature around 15 mK, a process that took  6 hours to 4.2 K and an additional 6 hours to base temperature. 
The same chip was cooled down from room temperature four times, with essentially identical results.  During each cooldown, the local thermometer was calibrated against a $\mathrm{RuO_2}$ resistor mounted to the mixing chamber.

A multi-mode optical fibre was used to shine light onto the sample from a red LED located outside the refrigerator. Below-bandgap light was found to have no effect.  With a power of 140 mW applied to the LED (80 mA at 1.7 V), $6\pm 2$ $\mu$V of optical power reached the bottom of the fibre, as estimated from the 15 mK rise from base temperature in the mixing chamber temperature during illumination. Geometrical considerations suggest an optical intensity of $120 \pm 50$ nW/mm$^2$ at the location of the sample. The data presented here are taken for the sample brought from room temperature without fibre illumination during the cooldown.

After arriving at base temperature, and prior to illumination, the longitudinal  ($\mathrm{R}_{xx}$) and Hall ($\mathrm{R}_{xy}$) magnetoresistance were measured at 30 mK using a lock-in amplifier at 53.4 Hz (Figs.~\ref{beforeaftertreatment}a,b). The bias current was kept low enough to ensure that it was not heating the sample, typically 100 nA below 100 mT and 2 nA at high field. 
The temperature (30 mK) was selected to provide a repeatable baseline for very low temperature FQH characteristics, cold enough that most states were fully or almost fully developed ($\mathrm{R}_{xx}\rightarrow 0$) but warm enough that we could return easily to this temperature without waiting many hours for thermal equilibration.

The carrier density determined from the Hall slope was 90\% of the nominal density, though variations of several percent were common from cooldown to cooldown; the Drude mobility was $\mu=1.45\pm0.05 \times 10^7$ cm$^2$/Vs. Fractional quantum Hall features in the $\nu=2-4$ range were only weakly developed; no reentrant integer quantum Hall states were visible (Figs.~\ref{beforeaftertreatment}a,b).\cite{FirstRIQHSObs,RIQHSinSLL}

Next, the sample was illuminated for 30 minutes, during which the chip carrier temperature (monitored by the local thermometer) warmed to 600 mK due to the optical power. The light was turned off, then the sample was annealed using the local heater  at a temperature between 2.2 K and 6.5 K for a period of time between 4 and 48 minutes, before cooling again to 30 mK.
Figs.~\ref{beforeaftertreatment}c,d show $\mathrm{R}_{xx}$ and $\mathrm{R}_{xy}$ after illumination and 20 minutes of annealing at 2.25 K. Compared to sample characteristics immediately after cooldown in the dark, carrier density and mobility have only slightly increased ($\mu=1.5\pm0.05 \times 10^7$ cm$^2$/Vs) but many more fractional and reentrant states are visible.

The improvement of FQH quality as a function of annealing time was gradual, taking tens of minutes or longer for annealing temperatures below 2.1 K, but less than 4 minutes above 2.5 K.  When annealing at 2.25 K, for example, FQH quality increased with annealing time for the first 10-15 minutes, after which the quality was saturated at its highest level.  Fig.~\ref{evolution} shows longitudinal and Hall resistances at 30 mK between filling factors $\nu$=2 and 3, for annealing times of 4, 8, and 20 minutes at 2.25 K.  Before each anneal step, the sample was reset by a 30 minute illumination.  After only a 4 minute anneal the FQH features are poorly formed, but after 8 minutes the features as they appear at 30 mK are close to their saturated form, as seen by comparing to the 20 minute data.

\begin{figure}[tp]
	\begin{center}
	\includegraphics{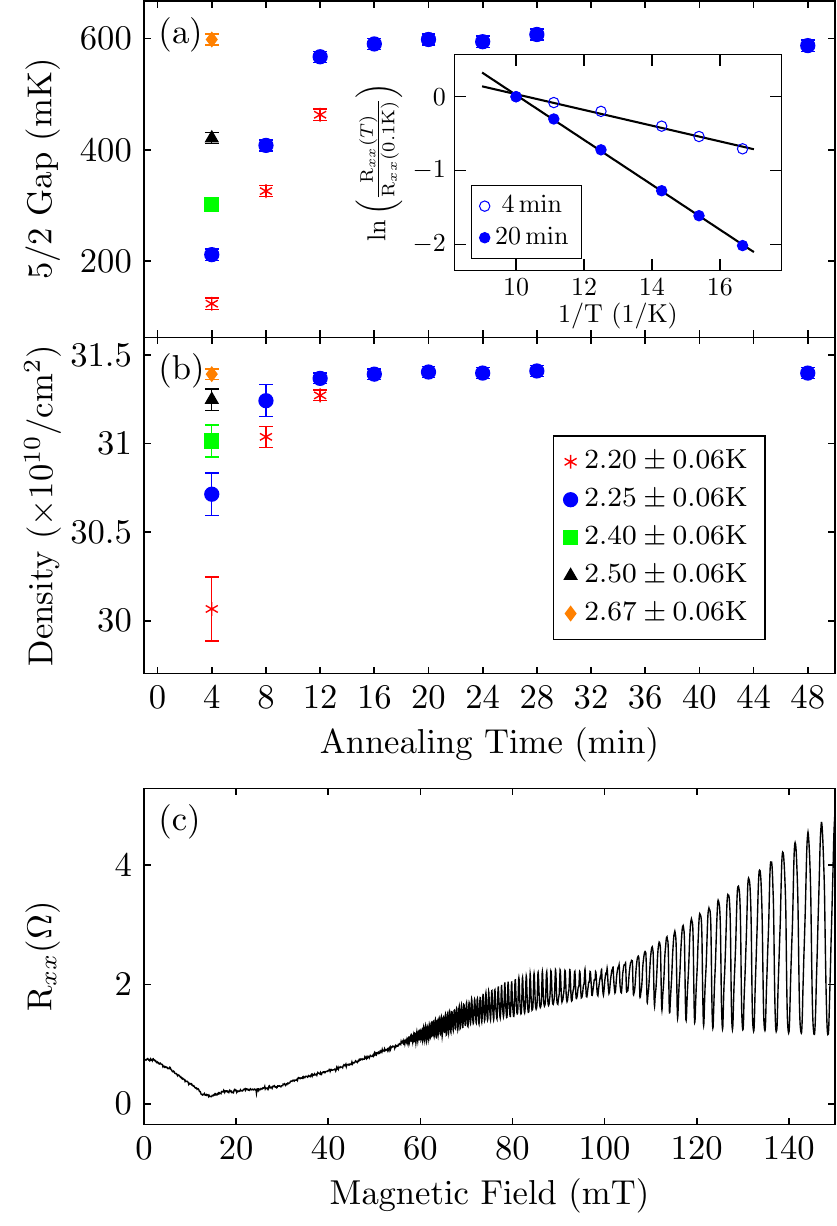}
	\caption{$\Delta_{5/2}$ (a) and carrier density (b) increase gradually with annealing time, at a rate that increases dramatically with annealing temperature. Before each data point, the sample was illuminated for 30 minutes.  Inset to (a) is example Arrhenius plot for two annealing times at 2.25 K. (c)  4 minutes of annealing at 2.20 K yields Shubnikov-de Haas oscillations with strong beats, suggesting inhomogeneous carrier density.
	\label{timedep}}
	\end{center}
\end{figure}

In order to quantify the improvement in sample quality with annealing temperature and time, the carrier density and activation energy gap of the $\nu=5/2$ FQH state, $\Delta_{5/2}$, were recorded for many different anneal parameters (Figs.~\ref{timedep}a,b). Carrier density, $n_s$, was extracted from the magnetic field location of the $\nu=5/2$ minimum by the relation $\nu=n_sh/eB$, where $h$ is the Planck's constant and $e$ is the elementary charge.  The activation energy gap was extracted by fitting the temperature dependence of the minimum in $\mathrm{R}_{xx}$ near $\nu=5/2$ to the Arrhenius formula, 
\begin{align}
    \mathrm{R}_{xx}(T) \propto \exp\left[\frac{-\Delta_{5/2}}{2k_BT}\right],
    \label{Arrh}
\end{align}
for Boltzmann constant $k_B$, see e.g.~Fig.~\ref{timedep}a(inset). Temperature dependence was measured between T=60 mK and T=100 mK, controlled by the mixing chamber heater and RuO$_2$ thermometer.   At lower temperatures the value of $\mathrm{R}_{xx}$ was too low to be measured accurately, while at higher temperatures $\mathrm{R}_{xx}$ no longer followed the exponential dependence of Eq.~\ref{Arrh}. After each temperature change, the resistance of the sample was monitored over time to ensure that thermal equilibration with the mixing chamber was complete. This took less than 2 minutes for T$<$70 mK, rising to 10 minutes for T=60 mK.

Comparing a 4 minute anneal at several different temperatures, it is interesting to note that the annealing speed increases dramatically as the anneal temperature increases from 2.2 to 2.5 K. Above 2.6 K, the process was so fast that full equilibration was obtained within 4 minutes, the shortest anneal time we performed. No degradation in any of the measured sample qualities was observed for anneals up to 6.5 K, the highest annealing temperature accessible in this experiment. The 2.25 K data in Fig.~\ref{timedep} also indicate that very long anneals do not degrade the $\nu=5/2$ gap, or change the  carrier density beyond its saturated value.  

\begin{figure}[t]
	\begin{center}
	\includegraphics{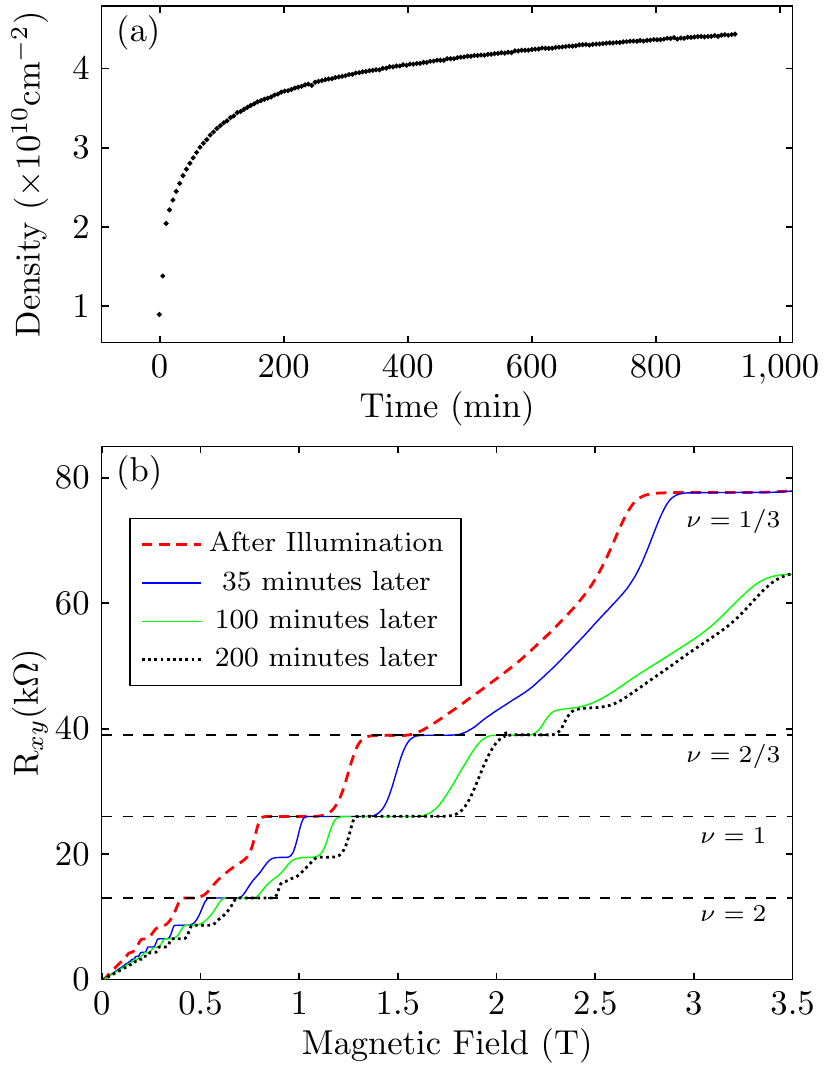}
	\end{center}
	\caption{(a)  After illumination is switched off, the density increases over time (T=30mK).  Illumination was switched off after the first point but before the second point on the graph. (b) Evolution of Hall traces after illumination at base temperature, with no subsequent annealing. The first trace (dashed line) begins 5 minutes after illumination was switched off. \label{shineonly}}
\end{figure}

Carrier density, $n_s$, equilibrates on a timescale similar to the evolution of $\Delta_{5/2}$.  On the other hand, the carrier density increased by only 3\% from 4 minutes to 20 minutes anneal time at 2.25 K, whereas $\Delta_{5/2}$ increased by nearly a factor of 3.  Apparently, the improvement in $\Delta_{5/2}$ for longer anneals is not simply the result of higher densities, and not the result of higher Drude mobility (the observed increase from 1.45 to 1.5$\times 10^7$cm$^2$/Vs is within the measurement error). 

We speculate that the higher reported $\Delta_{5/2}$ values may instead reflect improvements in the sample homogeneity.  The values for $\Delta_{5/2}$ reported in Fig.~3 are simply results of a fit to Eq.~\ref{Arrh}, and may not represent a homogeneous gap across the sample. Although the fits to  Eq.~\ref{Arrh} were always reasonable over the range 60-100 mK, the shape of the dip in $\mathrm{R}_{xx}(B)$ near $\nu=5/2$ was often asymmetric when annealing times were insufficient---not simply less deep as one might expect for a weakly-gapped state. Misshapen features in $\mathrm{R}_{xx}$ may indicate a spatially varying charge density, such that macroscopic measurements of resistivity represent  averages over many regions in the sample, each with slightly different filling factor. This would be consistent with the strong beats observed in Shubnikov de Haas oscillations for short annealing times (Fig.~\ref{timedep}c), which disappear for longer anneals (data not shown), thus  suggesting that longer annealing times enhance the density homogeneity.



A microscopic description of the process by which dopants equilibrate to screen potential fluctuations has not, to our knowledge, been worked out.  However, the extreme temperature dependence of the annealing rate may indicate a thermally activated diffusion process, where charges hop between localized sites in or near the doping well with a diffusion constant $D\propto T\cdot\exp[-\Delta_H/k_BT]$, where $\Delta_H$ is the energy barrier to hopping. Noting that, in Figs.~\ref{timedep}a,b, the values are similar for [4 minutes, 2.5 K] and [8 minutes, 2.25 K], and also for [4 minutes, 2.4 K] and [8 minutes, 2.2 K], we can estimate D(T=2.5 K)$\approx$ 2D(T=2.25 K), and D(T=2.4 K)$\approx$ 2D(T=2.2K), giving $\Delta_H$  between 13 and 16 K.

Although carrier density was within a few percent of its maximum value after only 4 minutes of annealing (Fig.~3b), the carrier density immediately after illumination was a factor of 30 lower.  Similar effects have previously been observed in GaAs heterostructures.\cite{DensityReduction}
In our experiment, the density remained very low (less than $4.4 \times 10^{10}$/cm$^2$) for many hours as long as the sample was kept below 100 mK.  The time dependence of the increase in density after illumination was monitored using the slope of the Hall resistance near zero field, and is shown in Fig.~\ref{shineonly}a.

Figure 4b shows four consecutive Hall resistance traces, taken after 30 minutes of illumination with no annealing.
The first trace was begun a few minutes after illumination was switched off (the sample had returned to less than 50 mK by this time), then subsequent traces were taken over the next three hours.
Note that $\nu=1$ occurs for $B\sim1$ T in Fig.~\ref{shineonly}, compared to 13 T where the $\nu=1$ would appear for annealed samples (c.f.~Figs.~1,2,3).
The Drude mobility, as determined by the low field $R_{xx}$, was $4\times 10^6$ cm$^2$/Vs for the last trace when density was $4.4 \times 10^{10}$/cm$^2$.
Several of the $\nu=1/3$ ladder of fractional states can be clearly seen, increasing in strength for the later traces.

In conclusion, we found that optimal FQH characteristics of a doping well two-dimensional electron gas can be obtained by illumination at low temperature, then annealing at temperatures between 2.2 K and 6.5 K. The time scale for annealing depends sensitively on the temperature, but too-long anneals do not appear to degrade FQH characteristics.  Illumination can also be used to \textit{reset} the sample to a well-defined initial state with very low density.
This technique can be used to compare FQH characteristics of quantum well samples grown under different MBE protocols, without having to rely on a precise cooldown recipe that is inherently difficult to control.

Experiments at UBC were supported by NSERC, CFI, and CIFAR. ES and SL are partially supported by the Quantum Matter Institute. The molecular beam epitaxy growth at Purdue is supported by the U.S. Department of Energy, Office of Basic Energy Sciences, Division of Materials Sciences and Engineering under Award DE-SC0006671.

\bibliography{main}

\end{document}